\documentclass[%
    aps,
    pra,
    reprint,
    10pt,
    superscriptaddress,
    amsfonts,
    amssymb,
    amsmath,
    preprintnumbers,
    noshowpacs,
    tightenlines,
    floats,
    notitlepage,
    final,
    letterpaper,
    oneside,
    twocolumn,
    footinbib,
    longbibliography,
    ]{revtex4-1}
\pdfoutput=1
\usepackage[utf8]{inputenc}
\usepackage[T1]{fontenc}
\usepackage[american]{babel}
\usepackage[british,cleanlook,printdayon]{isodate}
\usepackage[Symbolsmallscale]{upgreek}
\usepackage{mathtools}
\usepackage{isomath}
\usepackage[protrusion,tracking,kerning,spacing,final,babel]{microtype}
\usepackage{physics}
\usepackage[final]{hyperref}
\hypersetup{%
    colorlinks=true,
    citecolor=red,
    linkcolor=red,
    urlcolor=red,
    }%
\usepackage{cleveref}
\usepackage{orcidlink}

\newcommand{\rme}{\mathrm{e}}
\newcommand{\rmi}{\mathrm{i}}
\newcommand{\down}{\downarrow}
\newcommand{\up}{\uparrow}

\begin{document}

\title{Generalized cold-atom simulators for vacuum decay}

\author{Alexander~C.~Jenkins\,\orcidlink{0000-0003-1785-5841}}
\altaffiliation{Corresponding author}
\email{alex.jenkins@ucl.ac.uk}
\affiliation{Department of Physics and Astronomy, University College London, London WC1E 6BT, UK}

\author{Ian~G.~Moss\,\orcidlink{0000-0003-0155-9394}}
\affiliation{School of Mathematics, Statistics and Physics, Newcastle University, Newcastle upon Tyne NE1 7RU, UK}

\author{Thomas~P.~Billam\,\orcidlink{0000-0003-2456-6204}}
\affiliation{Joint Quantum Centre (JQC) Durham--Newcastle, School of Mathematics, Statistics and Physics, Newcastle University, Newcastle upon Tyne NE1 7RU, UK}

\author{Zoran Hadzibabic\,\orcidlink{0000-0002-0118-9285}}
\affiliation{Cavendish Laboratory, University of Cambridge, J.~J.~Thomson Avenue, Cambridge CB3 0HE, UK}

\author{Hiranya~V.~Peiris\,\orcidlink{0000-0002-2519-584X}}
\affiliation{Institute of Astronomy and Kavli Institute for Cosmology, University of Cambridge, Madingley Road, Cambridge CB3 0HA, UK}
\affiliation{The Oskar Klein Centre for Cosmoparticle Physics, Department of Physics, Stockholm University, AlbaNova, Stockholm SE-106 91, Sweden}

\author{Andrew~Pontzen\,\orcidlink{0000-0001-9546-3849}}
\affiliation{Institute for Computational Cosmology, Department of Physics, Durham University, South Road, Durham, DH1 3LE, UK}
\affiliation{Department of Physics and Astronomy, University College London, London WC1E 6BT, UK}

\begin{abstract}
    Cold-atom analog experiments are a promising new tool for studying relativistic vacuum decay, enabling one to empirically probe early-Universe theories in the laboratory.
    However, existing proposals place stringent requirements on the atomic scattering lengths that are challenging to realize experimentally.
    Here we eliminate these restrictions and show that \emph{any} stable mixture between two states of a bosonic isotope can be used as a faithful relativistic analog.
    This greatly expands the landscape of suitable experiments, and will expedite efforts to study vacuum decay with cold atoms.
\end{abstract}

\date{\today}
\maketitle

\textbf{\emph{Introduction.}}---Quantum fields can escape metastable `false vacuum' states by nucleating `true vacuum' bubbles~\cite{Coleman:1977py,Callan:1977pt,Coleman:1980aw,Linde:1981zj,Weinberg:2012pjx}.
This process of \emph{vacuum decay} plays a pivotal role in many aspects of cosmology, from the Universe's inflationary beginnings~\cite{Guth:2007ng,Aguirre:2007gy,Aguirre:2007an,Feeney:2010jj,Feeney:2010dd} to the present-day (meta)stability of the Higgs field~\cite{Ellis:2009tp,Degrassi:2012ry,Buttazzo:2013uya}.
The standard treatment of this problem relies on Euclidean (imaginary-time) calculations, leaving many key questions unanswered.
In particular, how does vacuum decay proceed in real time?
And what happens in situations where the symmetries of the Euclidean solutions are broken, e.g., nucleation of multiple bubbles~\cite{Pirvu:2021roq}?

There has been a recent surge of interest in tackling these questions using quantum analog experiments that simulate metastable relativistic fields~\cite{Opanchuk:2013lgn,Fialko:2014xba,Fialko:2016ggg,Braden:2017add,Billam:2018pvp,Braden:2019vsw,Billam:2020xna,Ng:2020pxk,Billam:2021qwt,Billam:2021nbc,Billam:2022ykl,Jenkins:2023eez}.
In this \emph{Letter}, we present the first such proposal that is both viable with current experiments and rigorously, quantitatively analogous to relativistic vacuum decay.
Our proposal uses ultracold atomic condensates to enable controlled real-time tests of early-Universe theories on a tabletop.
Similar technologies have been successfully used to study discontinuous transitions in nonrelativistic quantum fields~\cite{Struck:2013ixy,Campbell:2016soc,Trenkwalder:2016qpt,Qiu:2020kzm,Song:2021pyy,Cominotti:2022jrj}, including nonrelativistic thermal vacuum decay~\cite{Zenesini:2023afv}.

As we demonstrate below, our proposed analog system satisfies four requirements that are essential for delivering new cosmological insights~\footnote{With deviations that can be made perturbatively small by tuning the experimental parameters.}:
\begin{enumerate}
    \item The system possesses a degree of freedom with the same \emph{equations of motion} as a relativistic field.
    \item The \emph{quantum fluctuation statistics} in this degree of freedom match those of the same relativistic field.
    \item The system can be made \emph{homogeneous} in one or more dimensions, replicating the translation-invariance of a relativistic theory.
    \item The effective relativistic field can be put into a \emph{metastable state}, leading to bubble nucleation.
\end{enumerate}

In mixtures between two atomic hyperfine states, this level of analogy has previously only been established under `symmetric' conditions, with equal density and $s$-wave scattering length in each state, and zero scattering between atoms in different states~\cite{Fialko:2014xba,Fialko:2016ggg,Braden:2017add,Braden:2019vsw,Billam:2020xna,Billam:2021qwt,Jenkins:2023eez}.
These conditions are challenging to realize in practice.
While the scattering can be tuned using Feshbach resonances in an applied magnetic field~\cite{Chin:2010fesh}, there is insufficient freedom to simultaneously satisfy all the symmetric conditions.
As a result, previous studies have found only two discrete points in parameter space where these conditions hold (both in ${}^{41}\mathrm{K}$~\cite{Fialko:2014xba,Fialko:2016ggg,Jenkins:2023eez}), leaving no flexibility to accommodate other experimental requirements.

\begin{figure*}[t!]
    \centering
    \includegraphics{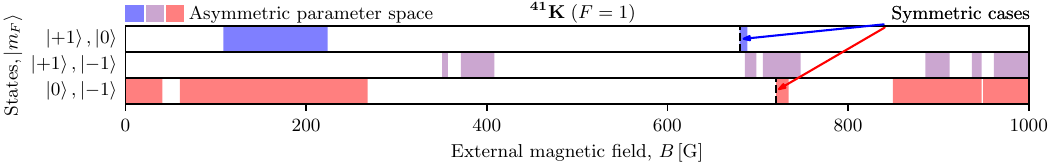}
    \caption{\label{fig:landscape-41K.pdf}
    Landscape of vacuum decay analogs in the $F=1$ hyperfine manifold of ${}^{41}\mathrm{K}$.
    The horizontal bars correspond to pairs of states $\ket{m_F}\in\{\ket{+1},\ket{0},\ket{-1}\}$.
    Shading indicates the viable parameter space for our asymmetric proposal (identified using scattering lengths from Ref.~\cite{Lysebo:2010fesh}), while dashed black lines show two points where the symmetric conditions hold~\cite{Fialko:2014xba,Fialko:2016ggg,Jenkins:2023eez}.
    (These lines are $\sim1000\times$ thicker than the actual symmetric regions.)
    As well as greatly expanding the parameter space for ${}^{41}\mathrm{K}$, our results enable analogs in isotopes with no symmetric options.}
\end{figure*}

Here we eliminate these restrictions, and show that \emph{any} stable mixture between two states of a bosonic isotope can be made to simulate relativistic vacuum decay by tuning their number densities.
While Ref.~\cite{Fialko:2016ggg} first studied such `asymmetric' systems in this context, there has until now been no theoretical justification for using them as vacuum decay analogs.
We establish this analogy rigorously, showing that asymmetric condensates can be made to exhibit all four conditions listed above, just as in the symmetric case.
By greatly expanding the landscape of experimental possibilities (see Fig.~\ref{fig:landscape-41K.pdf}), our results bring robust vacuum decay analogs within reach of current technologies.

\textbf{\emph{Analog system.}}---We consider a gas of two hyperfine states (`$\down$' and `$\up$') of some bosonic isotope.
At low temperatures, each forms a condensate described by a nonrelativistic quantum field $\hat{\psi}_i(\vb*x)=\sqrt{\hat{n}_i(\vb*x)}\,{\exp}(\rmi\hat{\phi}_i(\vb*x))$ ($i={\down,\up}$), whose modulus $\hat{n}_i(\vb*x)$ measures the density of atoms in state $i$, and whose phase $\hat{\phi}_i(\vb*x)$ captures coherent behavior.
We define the mean density $n=(n_\down+n_\up)/2$ and population imbalance $z=(n_\down-n_\up)/(n_\down+n_\up)$, which we treat as homogeneous.
In practice, the potential that traps the atoms generates an inhomogeneous density profile---however, using an optical box trap~\cite{Gaunt:2013box,Navon:2021mcf} ensures homogeneity everywhere except a small boundary region.

We couple the species with an electromagnetic field of frequency $\omega=\omega_0+\delta$, where $\omega_0=(E_\up-E_\down)/\hbar$ is the resonant frequency (with $E_i$ the energies of the states), and $\delta$ is the detuning away from resonance.
This causes interstate transitions at a variable rate $\Omega(t)$, set by the coupling amplitude.
The Hamiltonian is
    \begin{align}
    \begin{split}
    \label{eq:hamiltonian}
        \hat{H}(t)=&\int_V\dd{\vb*x}\bigg\{-\hat{\psi}_\down^\dagger\frac{\hbar^2\laplacian}{2m}\hat{\psi}_\down-\hat{\psi}_\up^\dagger\frac{\hbar^2\laplacian}{2m}\hat{\psi}_\up\\
        &-\mu(\hat{\psi}_\down^\dagger\hat{\psi}_\down+\hat{\psi}_\up^\dagger\hat{\psi}_\up)-\frac{\hbar\Omega(t)}{2}(\hat{\psi}_\down^\dagger\hat{\psi}_\up+\hat{\psi}_\up^\dagger\hat{\psi}_\down)\\
        &-\frac{\hbar\delta}{2}(\hat{\psi}_\down^\dagger\hat{\psi}_\down-\hat{\psi}_\up^\dagger\hat{\psi}_\up)+\sum_{i,j}g_{ij}\hat{\psi}_i^\dagger\hat{\psi}_j^\dagger\hat{\psi}_i\hat{\psi}_j\bigg\},
    \end{split}
    \end{align}
    with $m$ the atomic mass, $\mu$ the chemical potential~\footnote{This is the chemical potential for $N=N_\down+N_\up$, while $\mu_\down,\mu_\up$ are associated with $N_\down$ and $N_\up$ individually.}, and $g_{ij}$ describing contact interactions between species $i$ and $j$, which we parameterize as
    \begin{equation}
        g=\frac{g_{\down\down}+g_{\up\up}}{2},\quad\Delta=\frac{g_{\up\up}-g_{\down\down}}{2},\quad\kappa=\frac{g_{\down\down}+g_{\up\up}-2g_{\down\up}}{2}.
    \end{equation}
We require repulsive self-interactions, $g_{\down\down},g_{\up\up}>0$, while $g_{\down\up}=g_{\up\down}$ can have either sign.
The condition for miscibility (i.e., a stable homogeneous mean field) is $g_{\down\down}g_{\up\up}>g_{\down\up}^2$~\cite{Pethick:2008bec}, or equivalently $g^2-\Delta^2>(g-\kappa)^2$.

\begin{figure}[t!]
    \centering
    \includegraphics{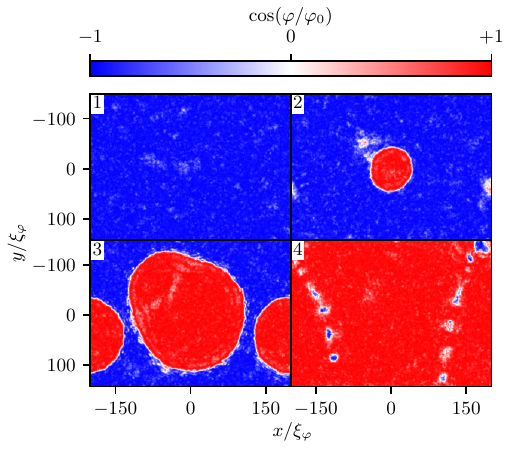}
    \caption{\label{fig:bubble-nucleation}
    Lattice simulation of a ${}^{39}\mathrm{K}$ analog, with time increasing from left to right and top to bottom.
    The metastable `false vacuum' (blue) spontaneously decays via expanding bubbles of `true vacuum' (red).
    We show a 2D analog here for illustration, while our main numerical results are in 1D.}
\end{figure}

The relative phase in this system, $\phi_\down-\phi_\up$, behaves like a relativistic scalar field, with a potential determined by the coupling $\Omega(t)$.
To recover a false-vacuum potential, this coupling must contain a constant piece and an oscillating piece.
The constant piece generates a cosine potential for the relative phase, due to the cost of putting the species out of phase with each other.
This does not allow vacuum decay, as the antiphase `false vacuum' is then unstable rather than metastable.
As first pointed out by \citet{Fialko:2014xba} however, this state can be made metastable by modulating the coupling.
This generates an effective potential barrier, allowing for discontinuous transitions out of the false vacuum (as illustrated in Fig.~\ref{fig:bubble-nucleation}).

We therefore set $\hbar\Omega(t)=2\epsilon n\sqrt{\kappa^2-\Delta^2}+\lambda\hbar\nu\sqrt{2\epsilon}\cos\nu t$, where $\nu\gg\kappa n/\hbar$ is the modulation frequency, and $\epsilon,\lambda$ control the amplitudes of the constant and oscillatory terms~\footnote{Note that $\Omega(t)$ can change sign, modulating the phase as well as the amplitude of the coupling.}.
Inserting this into Eq.~\eqref{eq:hamiltonian}, we use the formalism of Ref.~\cite{Goldman:2014xja} to calculate an effective Hamiltonian, valid for $\epsilon\ll1$ and on timescales $\gg2\uppi/\nu$~\footnote{See Supplemental Material for details regarding the effective Hamiltonian, vacuum states, relativistic equations of motion, vacuum fluctuations, and Euclidean action.}.
The false-vacuum barrier stems from quadratic terms in the modulation, which are $\order*{\epsilon}$.
The constant part of the coupling is also $\order*{\epsilon}$, to ensure the energy difference between vacua is comparable to the barrier height; the exact relationship depends on the $\order*{1}$ parameter $\lambda$, where $\lambda\ge1$ is required for metastability, and larger values give longer-lived false vacua.

\textbf{\emph{Effective relativistic theory.}}---Our key result concerns the phase fields $\hat{\phi}_i(\vb*x)$, which we rewrite as
    \begin{align}
    \begin{split}
    \label{eq:fields}
        \hat{\vartheta}(\vb*x)&=\sqrt{\frac{\hbar^2n}{2m}}\qty((1+z)\hat{\phi}_\down(\vb*x)+(1-z)\hat{\phi}_\up(\vb*x)),\\
        \hat{\varphi}(\vb*x)&=\sqrt{\frac{\hbar^2n}{2m}}\sqrt{1-z^2}(\hat{\phi}_\down(\vb*x)-\hat{\phi}_\up(\vb*x)).
    \end{split}
    \end{align}
In situations where the detuning $\delta$, density fluctuations $n_i-\ev{n_i}$, and field gradients $\laplacian{\vartheta},\laplacian{\varphi}$ are all $\order*{\epsilon}$, one can eliminate the densities to obtain coupled equations for $\vartheta,\varphi$~\cite{Note4}.
For the appropriate population imbalance,
    \begin{equation}
    \label{eq:decoupling}
        z=\frac{\Delta}{\kappa}+\order*{\epsilon},
    \end{equation}
    the equations decouple to give
    \begin{align}
    \begin{split}
    \label{eq:rel-eoms}
        0&=(c_\vartheta^{-2}\partial_t^2-\laplacian)\vartheta,\\
        0&=(c_\varphi^{-2}\partial_t^2-\laplacian)\varphi+U'(\varphi).
    \end{split}
    \end{align}
These are \emph{relativistic} equations of motion for a massless scalar $\vartheta$ and a self-interacting scalar $\varphi$ with potential
    \begin{equation}
    \label{eq:potential}
        U(\varphi)=m_0^2\varphi_0^2\frac{c_\varphi^2}{\hbar^2}\qty[1-\cos(\varphi/\varphi_0)+\frac{\lambda^2}{2}\sin^2(\varphi/\varphi_0)],
    \end{equation}
    where we define $m_0=m\sqrt{4\epsilon\kappa^2/(\kappa^2-\Delta^2)}$ and $\varphi_0=\sqrt{(\kappa^2-\Delta^2)\hbar^2n/(2m\kappa^2)}$.
This potential contains true vacua at $\varphi_\mathrm{tv}/\varphi_0=0\pmod{2\uppi}$, and, for $\lambda>1$, false vacua at $\varphi_\mathrm{fv}/\varphi_0=\uppi\pmod{2\uppi}$, allowing one to observe vacuum decay by preparing the system in a false vacuum.
The decoupling between $\vartheta$ and $\varphi$ is crucial for simulating a relativistic system, as the general coupled equations cannot be derived from any relativistic Lagrangian.

The fields~\eqref{eq:fields} each live on a flat Minkowski spacetime with its own `speed of light' (i.e., phonon sound speed),
    \begin{equation}
    \label{eq:sound-speeds}
        c_\vartheta^2=\frac{n}{m}\qty(2g-\kappa-\frac{\Delta^2}{\kappa}),\quad c_\varphi^2=\frac{n}{m}\qty(\kappa-\frac{\Delta^2}{\kappa}),
    \end{equation}
    where miscibility implies $c_\varphi^2\ge c_\vartheta^2>0$.
These define two healing lengths, $\xi_i=\hbar/(\sqrt{2}mc_i)$, below which each field becomes nonrelativistic, thus determining the regime of validity of the analogy.

References~\cite{Fischer:2004bf,Visser:2004qp,Visser:2005ss,Weinfurtner:2006wt} previously investigated this `bi-metric' structure in Bose-Bose mixtures, deriving a condition analogous to Eq.~\eqref{eq:decoupling} for linear perturbations around a fixed background when $g_{\down\down}=g_{\up\up}$ and/or $g_{\down\up}=0$.
Here we generalize to any $g_{ij}$, greatly expanding the parameter space, and derive a fully nonlinear relativistic theory for the phase fields, which is crucial for simulating nonperturbative phenomena like vacuum decay.

Our decoupling condition~\eqref{eq:decoupling} is equivalent to matching the chemical potentials of the two species, $\mu_\down=\mu_\up$.
Intuitively, this is because each chemical potential sets the mean velocity of the corresponding phase, so Eq.~\eqref{eq:decoupling} ensures that $\ev*{\dot{\varphi}}=0$ in the false vacuum.
This also suppresses density fluctuations, as required for pseudorelativistic phase dynamics.

As well as the equations of motion~\eqref{eq:rel-eoms}, we have verified that the analogy holds in terms of the vacuum fluctuations~\cite{Note4}, which is crucial for faithfully simulating the decay rate~\cite{Jenkins:2023eez}.
Expanding the effective Hamiltonian to quadratic order in the fluctuations, we find that it decouples into two independent sectors, $\hat{H}_\mathrm{eff}\simeq\hat{H}_\vartheta+\hat{H}_\varphi$, if and only if Eq.~\eqref{eq:decoupling} holds.
Diagonalizing each sector we find that, on scales much larger than their respective healing lengths ($\xi_ik\ll1$, with $k$ the wavenumber), the dispersion relations match those of a massless relativistic scalar $\vartheta$ and a self-interacting scalar $\varphi$ with potential given in Eq.~\eqref{eq:potential}, with the sound speeds given by Eq.~\eqref{eq:sound-speeds},
    \begin{equation}
        \omega_{\vartheta}(k)\simeq c_\vartheta k,\quad\omega_{\varphi}(k)\simeq\sqrt{c_\varphi^2k^2+\frac{m_\mathrm{fv}^2c_\varphi^4}{\hbar^2}},
    \end{equation}
    where $\omega_\vartheta(k),\omega_\varphi(k)$ are the frequencies of modes with wavenumber $k$, and $m_\mathrm{fv}=\sqrt{\lambda^2-1}\,m_0$.
This nontrivial check confirms all the key features of the theory.

\textbf{\emph{Experimental proposal.}}---Our results open an extensive new landscape of experimental possibilities.
As well as greatly expanding the ${}^{41}\mathrm{K}$ parameter space beyond the two symmetric cases considered in Refs.~\cite{Fialko:2014xba,Fialko:2016ggg,Jenkins:2023eez} (see Fig.~\ref{fig:landscape-41K.pdf}), we can now consider other isotopes, where no symmetric cases exist.
As an illustrative example, we present a feasible set of parameters for ${}^{39}\mathrm{K}$.
We consider the $F=1$ hyperfine states $\ket{\down}\equiv\ket{m_F=0}$ and $\ket{\up}\equiv\ket{m_F=-1}$ (with $F,m_F$ the total and projected angular momenta).
These are miscible in a magnetic field $B\approx57\text{--}59\,\mathrm{G}$, and have been used to study `quantum droplets' just outside the miscible regime~\cite{Cabrera:2018qld,Semeghini:2018sbq,Cheiney:2018sol,Ferioli:2019col}.

While the asymmetric analogy is valid in any number of dimensions, we focus here on the 1D case, with the atoms tightly confined in the transverse directions.
This offers several practical advantages, including the option of running arrays of traps in parallel, or enforcing periodicity using a ring trap.
Bubble nucleation is parametrically faster in 1D~\cite{Note4}, allowing one to probe a broader range of decay rates within each run.
A 2D setup is also possible (shown in Fig.~\ref{fig:bubble-nucleation} for illustration), and would allow one to investigate phenomena such as domain walls between distinct true vacua.
A 3D setup would carry technical complications associated with levitating both species against gravity~\cite{Navon:2021mcf}.

We summarize our proposed parameters in Table~\ref{tab:parameters}.
We choose these to maximize the energy scale $mc_\varphi^2$, thereby suppressing the influence of thermal fluctuations.
This results in a population imbalance $z\approx0.7$, which differs strongly from the symmetric case $z=0$.
To scan over decay rates, we vary the number of atoms inversely to the strength of transverse confinement, holding $mc_\varphi^2$ constant.
This lets one vary the dimensionless density $\bar{n}_\varphi=\xi_\varphi n$, which controls the fluctuation amplitudes, while leaving all other parameters of the relativistic theory fixed~\cite{Jenkins:2023eez}.
Since the rate scales like $\log\Gamma\sim-\bar{n}_\varphi$, even a small range of $\bar{n}_\varphi$ is sufficient to probe a broad range of scenarios.

One can initialize the false vacuum using a strong, constant coupling $\Omega\gg mc_\varphi^2/\hbar$, so the Hamiltonian becomes
    \begin{equation}
        \hat{H}\simeq-\frac{\hbar}{2}\mqty(\ket{\down} & \ket{\up})\mqty(\delta & \Omega \\ \Omega & -\delta)\mqty(\bra{\down} \\ \bra{\up})
    \end{equation}
    (temporarily ignoring inhomogeneous modes).
The resulting energy eigenstates have the species either in phase or in antiphase, corresponding to the true and false vacua, with the population imbalance in the latter given by
    \begin{equation}
        z_\mathrm{fv}=\frac{\braket{\mathrm{fv}}{\down}-\braket{\mathrm{fv}}{\up}}{\braket{\mathrm{fv}}{\down}+\braket{\mathrm{fv}}{\up}}=-\frac{\delta}{\sqrt{\delta^2+\Omega^2}}.
    \end{equation}
Starting from a pure-$\ket{\up}$ condensate with large positive detuning (or pure-$\ket{\down}$ with large negative detuning), one can thus initialize a false vacuum with arbitrary $z$ by adiabatically varying $\delta$.
We then replace the strong, constant coupling with the weak, modulated coupling required for pseudorelativistic dynamics.
After allowing the system to evolve for some time, $\varphi$ is converted via a radio-frequency pulse into a density contrast between true- and false-vacuum regions, and imaged with an efficiency of $\sqrt{1-z^2}\approx70\%$.
Analyzing snapshots from many runs then allows one to extract quantities like the decay rate.

\textbf{\emph{Lattice simulations.}}---We test our predictions with semiclassical lattice simulations, using the time-dependent Hamiltonian~\eqref{eq:hamiltonian} to evolve $\psi_\down,\psi_\up$.
We model the quantum nature of the system by including vacuum fluctuations in the initial state, which we evolve in real time by numerically integrating the classical equations of motion.
Repeating this for an ensemble of initial states gives a semiclassical approximation of the system's behavior, known as the `truncated Wigner approximation'~\cite{Blakie:2008vka,Braden:2018tky}.

These simulations provide detailed dynamical information, complementing insights from Euclidean calculations, and revealing a wealth of new observables such as bubble clustering~\cite{Pirvu:2021roq}, precursors~\cite{Pirvu:2023plk}, and time-dependent decay rates~\cite{Batini:2023zpi,Pirvu:2024ova,Pirvu:2024nbe} that are otherwise inaccessible, but could have important theoretical and observational implications.
Both the real-time (simulation-based) and imaginary-time (Euclidean) approaches make assumptions and approximations~\cite{Coleman:1977py,Callan:1977pt,Blakie:2008vka} that ultimately must be empirically tested and calibrated against experiments.
This comparison between simulations, Euclidean methods, and our proposed analog experiments has the potential to advance our understanding of vacuum decay, with wide-ranging impact across cosmology and high-energy physics.

\begin{table}[t!]
    \centering
    \begin{tabular}{l l}
        \hline\hline
        Parameter & Value \\
        \hline
        Atomic isotope & potassium-39 (${}^{39}\mathrm{K}$) \\
        Atomic mass & $m=38.96\,\mathrm{u}=6.470\times10^{-26}\,\mathrm{kg}$ \\
        Hyperfine states & $\ket{\down}\equiv\ket{F=1,m_F=0}$ \\
        & $\ket{\up}\equiv\ket{F=1,m_F=-1}$ \\
        Magnetic field & $B=58.50\,\mathrm{G}$ \\
        Scattering lengths (3D) & $a_{\down\down}=31.85\,a_0=1.686\,\mathrm{nm}$ \\
        & $a_{\up\up}=446.2\,a_0=23.61\,\mathrm{nm}$ \\
        & $a_{\down\up}=-51.84\,a_0=-2.743\,\mathrm{nm}$ \\
        Population imbalance & $z=0.7159$ \\
        Healing lengths & $\xi_\vartheta=1.797\times10^4\,a_0=1.141\,\upmu\mathrm{m}$ \\
        & $\xi_\varphi=9.449\times10^3\,a_0=0.600\,\upmu\mathrm{m}$ \\
        Sound speeds & $c_\vartheta=1.010\,\mathrm{mm}\,\mathrm{s}^{-1}$ \\
        & $c_\varphi=1.921\,\mathrm{mm}\,\mathrm{s}^{-1}$ \\
        Energy scales & $mc_\vartheta^2=0.412\,\mathrm{peV}=4.78\,k_\mathrm{B}\,\mathrm{nK}$ \\
        & $mc_\varphi^2=1.490\,\mathrm{peV}=17.29\,k_\mathrm{B}\,\mathrm{nK}$ \\
        Box trap length & $L=400\,\xi_\varphi=240\,\upmu\mathrm{m}$ \\
        Sound-crossing time & $L/c_\varphi=124.9\,\mathrm{ms}$ \\
        Total number of atoms & $8000\le N\le32000$ \\
        Density per species & $16.67\,\upmu\mathrm{m}^{-1}\le n\le66.67\,\upmu\mathrm{m}^{-1}$ \\
        Dimensionless density & $10\le\bar{n}_\varphi\le40$ \\
        Transverse trap frequency & $0.356\,\mathrm{kHz}\le\omega_\bot/2\uppi\le1.43\,\mathrm{kHz}$ \\
        Mean Rabi frequency & $\Omega_0=16.04\,\mathrm{Hz}$ \\
        Rabi coupling parameter & $\epsilon=2.5\times10^{-3}$ \\
        Modulation amplitude & $\lambda=\sqrt{2}$ \\
        False vacuum mass & $m_\mathrm{fv}=\sqrt{\lambda^2-1}\,m_0=0.1425\,m$ \\
        \hline\hline
    \end{tabular}
    \caption{\label{tab:parameters}%
    Fiducial 1D experimental parameters.
    Here $\mathrm{u}=1.661\times10^{-27}\,\mathrm{kg}$ and $a_0=5.292\times10^{-2}\,\mathrm{nm}$.
    The dimensionless density $\bar{n}_\varphi=\xi_\varphi n$ is set by varying the number of atoms $N$ and the transverse trapping frequency $\omega_\bot$.}
\end{table}

We use these simulations here to test the effective relativistic theory described above.
We simulate a ${}^{39}\mathrm{K}$ mixture in a periodic 1D box, with parameters from Table~\ref{tab:parameters}, using an eighth-order symplectic pseudospectral code.
(See Ref.~\cite{Jenkins:2023eez} for details and convergence tests; we obtain the same level of convergence here.)
In each simulation we set $z$ by minimizing $\mu$ in the homogeneous false vacuum, with $\delta=0$.
This corresponds to the decoupling value~\eqref{eq:decoupling} with an $\order*{\epsilon}$ correction.
We then draw random fluctuations in $\vartheta,\varphi$ and their conjugate momenta, and use our chosen $z$ to convert these into fluctuations in $\psi_\down,\psi_\up$.

\begin{figure}[t!]
    \centering
    \includegraphics{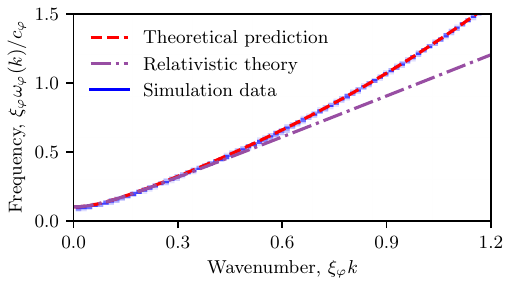}
    \caption{\label{fig:dispersion-relation}
    Dispersion relationship for $\varphi$ in the false vacuum.
    The blue heatmap shows the average of 32 simulations with parameters from Table~\ref{tab:parameters} (with $\bar{n}_\varphi=100$, so no bubbles nucleate).
    This dispersion relation matches that of a relativistic scalar of mass $m_\mathrm{fv}$ (purple) on large scales but becomes nonrelativistic on small scales, agreeing with our theoretical prediction (red).}
\end{figure}

We perform several tests, all confirming that $\vartheta,\varphi$ behave like decoupled relativistic fields if and only if Eq.~\eqref{eq:decoupling} holds:
\begin{enumerate}
    \item We test for decoupling by initializing $\varphi$ without fluctuations, while initializing $\vartheta$ as normal; given perfect decoupling, $\varphi$ should remain constant.
    For $z\approx0.7$, fluctuations in $\varphi$ remain small, reaching a maximum of $\approx0.1\,\%$ of the mean-field after one sound-crossing time.
    We interpret this as a weak coupling due to, e.g., neglected $\order*{\epsilon^2}$ terms.
    For $z=0$ the $\varphi$ fluctuations grow rapidly to equilibrate with those of $\vartheta$, demonstrating strong coupling.
    \item We test the dynamics of fluctuations (e.g., the sound speeds and false vacuum mass) by computing dispersion relations for $\vartheta,\varphi$.
    For $z\approx0.7$ we find excellent agreement with our predictions (see Fig.~\ref{fig:dispersion-relation}).
    For $z=0$ we find two branches in the dispersion relation of each field, reflecting the coupling between them, as well as modified sound speeds (as expected).
    \item We test whether the nonlinear dynamics of $\varphi$ are effectively relativistic by computing the Noether charges of the relativistic theory~\cite{Jenkins:2023eez}, with any time-variation interpreted as departures from relativistic behavior.
    We expect some variation due to nonrelativistic effects on small scales, as well as neglected renormalization corrections~\cite{Braden:2022odm}.
    However, we find that the charges are conserved just as well as in the symmetric case~\cite{Jenkins:2023eez}, with variation converging to zero as $\bar{n}_\varphi\to\infty$.
    This demonstrates a well-defined relativistic regime for the asymmetric analog.
\end{enumerate}

We perform these tests initially with large densities, $\bar{n}_\varphi\ge100$.
This suppresses the vacuum fluctuations, so that our perturbative calculations (e.g., treating density fluctuations as $\order*{\epsilon}$) give a good approximation.
For experimentally-accessible densities, $\bar{n}_\varphi\lesssim40$, we expect backreaction from fluctuations to become significant, leading to deviations from our predictions.
In particular, this should modify the critical $z$, so that Eq.~\eqref{eq:decoupling} no longer gives a long-lived metastable state.
We verify this numerically, finding that simulations with $\bar{n}_\varphi\lesssim40$ decay rapidly due to a homogeneous initial velocity in $\varphi$ that carries it over the barrier.

In principle one could account for backreaction analytically, calculating the renormalized critical value of $z$.
We postpone this for future work.
A simpler solution is to introduce a small detuning $\delta$ that compensates for the error in $z$, such that the initial mean velocity of $\varphi$ vanishes.
Adopting this prescription, our simulations exhibit precisely the expected behavior, passing all the tests described above.
This could be implemented experimentally using ensembles of runs with different detunings, selecting the ensemble with the longest-lived metastable state.

\textbf{\emph{Summary and outlook.}}---Cold-atom analog experiments raise the possibility of direct empirical tests of relativistic vacuum decay.
We have extended this relativistic analogy to a much larger class of cold-atom systems, representing the first proposal for analog vacuum decay that is both rigorously analogous to early-Universe theories \emph{and} experimentally viable with current capabilities.
These experiments have enormous discovery potential, allowing one to test decay rate predictions (potentially resolving or confirming the discrepancy between Euclidean calculations and lattice simulations~\cite{Braden:2018tky,Braden:2022odm}), as well as probing new phenomena, such as bubble clustering~\cite{Pirvu:2021roq} and dynamical precursors~\cite{Pirvu:2023plk}, with far-reaching potential implications for our understanding of the early Universe.

\textbf{\emph{Acknowledgments.}}---We thank Jonathan Braden, Christoph Eigen, Thomas Flynn, Matthew Johnson, Konstantinos Konstantinou, Dalila Pîrvu, Tanish Satoor, Silke Weinfurtner, and Paul Wong for valuable discussions.
This work was supported by the Science and Technology Facilities Council (STFC) [grant ST/T000708/1], the UK Quantum Technologies for Fundamental Physics programme [grants ST/T00584X/1, ST/T005904/1, ST/T006056/1, and ST/W006162/1], and the European Research Council (ERC) [UniFlat], and was partly enabled by the UCL Cosmoparticle Initiative.
ZH acknowledges support from the Royal Society Wolfson Fellowship.
This research was supported in part by Perimeter Institute for Theoretical Physics.
Research at Perimeter Institute is supported in part by the Government of Canada through the Department of Innovation, Science, and Economic Development Canada and by the Province of Ontario through the Ministry of Colleges and Universities.
We acknowledge the use of the Python packages NumPy~\cite{Harris:2020xlr}, SciPy~\cite{Virtanen:2019joe}, and Matplotlib~\cite{Hunter:2007ouj}.
The data that support the findings of this study are available from the corresponding author, ACJ, under reasonable request.

\textbf{\emph{Author contributions.}}---Based on the CRediT (Contribution Roles Taxonomy) system.
\textbf{ACJ}: methodology; software; formal analysis; investigation; data curation; interpretation and validation;  visualization; writing (original draft).
\textbf{IGM}: methodology; formal analysis; investigation; interpretation and validation; writing (review).
\textbf{TPB}: methodology; interpretation and validation; writing (review).
\textbf{ZH}: conceptualization; interpretation and validation; writing (review).
\textbf{HVP}: interpretation and validation; writing (review); project administration.
\textbf{AP}: interpretation and validation; writing (review).

\bibliography{fvd-39K}
\clearpage
\appendix
\section*{SUPPLEMENTAL MATERIAL}
\renewcommand\theequation{S\arabic{equation}}
\setcounter{equation}{0}

\subsection*{Time-averaged Hamiltonian}

The Hamiltonian can be split into a constant piece and an oscillating perturbation due to the modulation of the coupling,
    \begin{align}
    \begin{split}
        \hat{H}(t)&=\hat{H}_0-\lambda\hbar\nu\sqrt{2\epsilon}\,\hat{X}\cos(\nu t),\\
        \hat{X}&=\frac{1}{2}\int_V\dd{\vb*x}(\hat{\psi}^\dagger_\down\hat{\psi}_\up+\hat{\psi}^\dagger_\up\hat{\psi}_\down).
    \end{split}
    \end{align}
Using the formalism developed in Ref.~\cite{Goldman:2014xja}, we obtain an effective time-averaged Hamiltonian that governs the dynamics of modes with frequencies $\ll\nu$.
The small parameter $\epsilon$ associated with the coupling allows us to write this perturbatively as
    \begin{equation}
        \hat{H}_\mathrm{eff}=\hat{H}_0+\frac{\epsilon\lambda^2}{2}\comm*{\comm*{\hat{X}}{\hat{H}_0}}{\hat{X}}+\order*{\epsilon^2}.
    \end{equation}
Evaluating the commutators, this gives (to linear order in $\epsilon$)
    \begin{align}
    \begin{split}
    \label{eq:effective-hamiltonian}
        \hat{H}_\mathrm{eff}=&\int_V\dd{\vb*x}\bigg\{-\hat{\psi}_\down^\dagger\frac{\hbar^2\laplacian}{2m}\hat{\psi}_\down-\hat{\psi}_\up^\dagger\frac{\hbar^2\laplacian}{2m}\hat{\psi}_\up\\
        &-\mu(\hat{\psi}_\down^\dagger\hat{\psi}_\down+\hat{\psi}_\up^\dagger\hat{\psi}_\up)-\epsilon n\sqrt{\kappa^2-\Delta^2}(\hat{\psi}_\down^\dagger\hat{\psi}_\up+\hat{\psi}_\up^\dagger\hat{\psi}_\down)\\
        &-\frac{\hbar\delta}{2}\qty(1-\frac{\epsilon\lambda^2}{2})(\hat{\psi}_\down^\dagger\hat{\psi}_\down-\hat{\psi}_\up^\dagger\hat{\psi}_\up)\\
        &+\frac{1}{2}\qty(g-\Delta-\frac{\epsilon\lambda^2}{2}(\kappa-\Delta))\hat{\psi}_\down^\dagger\hat{\psi}_\down^\dagger\hat{\psi}_\down\hat{\psi}_\down\\
        &+\frac{1}{2}\qty(g+\Delta-\frac{\epsilon\lambda^2}{2}(\kappa+\Delta))\hat{\psi}_\up^\dagger\hat{\psi}_\up^\dagger\hat{\psi}_\up\hat{\psi}_\up\\
        &+\qty(g-\kappa\qty(1-\epsilon\lambda^2))\hat{\psi}_\down^\dagger\hat{\psi}_\down\hat{\psi}_\up^\dagger\hat{\psi}_\up\\
        &-\frac{\epsilon\lambda^2}{4}\kappa(\hat{\psi}_\down^\dagger\hat{\psi}_\down^\dagger\hat{\psi}_\up\hat{\psi}_\up+\hat{\psi}_\up^\dagger\hat{\psi}_\up^\dagger\hat{\psi}_\down\hat{\psi}_\down)\bigg\}.
    \end{split}
    \end{align}

As discussed in Refs.~\cite{Braden:2017add,Braden:2019vsw}, this time-averaging neglects the presence of parametric resonances driven by the modulation, which lead to an instability in modes with frequencies of $\order*{\nu}$.
In the experimental context, we expect to be able to damp this instability by making the modulation frequency sufficiently large.

\subsection*{Identifying the false vacuum state}

We consider first the dynamics of the zero mode of the system, with the aim of finding stationary mean-field solutions that we will identify with the true and false vacua of the effective relativistic theory.
Neglecting inhomogeneities, there are only two dynamical degrees of freedom: the relative phase $\varphi$, and the population imbalance $z$.
(In a homogeneous system the total density $n$ is constant, and the total phase $\vartheta$ is pure gauge.)
The equations of motion under the time-averaged Hamiltonian~\eqref{eq:effective-hamiltonian} are
    \begin{align}
    \begin{split}
    \label{eq:mean-field-equations-of-motion}
        \frac{\dot{\varphi}}{\varphi_0}&=-\frac{\Omega_0z}{\sqrt{1-z^2}}\cos(\varphi/\varphi_0)+(2-\epsilon\lambda^2)\qty(\frac{\Delta n}{\hbar}+\frac{\delta}{2})\\
        &-\frac{\kappa nz}{\hbar}\qty[2-\epsilon\lambda^2(3-\cos(2\varphi/\varphi_0))],\\
        \dot{z}&=\Omega_0\sqrt{1-z^2}\sin(\varphi/\varphi_0)+\epsilon\lambda^2\frac{\kappa n}{\hbar}(1-z^2)\sin\qty(2\varphi/\varphi_0),
    \end{split}
    \end{align}
    where $\Omega_0=(2\epsilon n/\hbar)\sqrt{\kappa^2-\Delta^2}$ is the constant part of the interspecies coupling.
These equations admit two stationary solutions, which we identify with the true and false vacua of the effective relativistic theory,
    \begin{equation}
        \varphi/\varphi_0=0\text{ or }\uppi,\quad z=\frac{\Delta}{\kappa}+\order*{\epsilon},
    \end{equation}
    i.e., the species are either in phase or in antiphase, with a population imbalance matching that in Eq.~\eqref{eq:decoupling}.
(The $\order*{\epsilon}$ term differs slightly between the two solutions, but as discussed above this correction has little to no effect on the dynamics.)

\begin{figure*}[t!]
    \centering
    \includegraphics{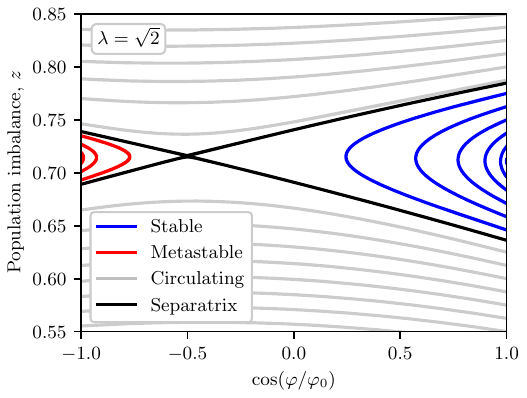}
    \includegraphics{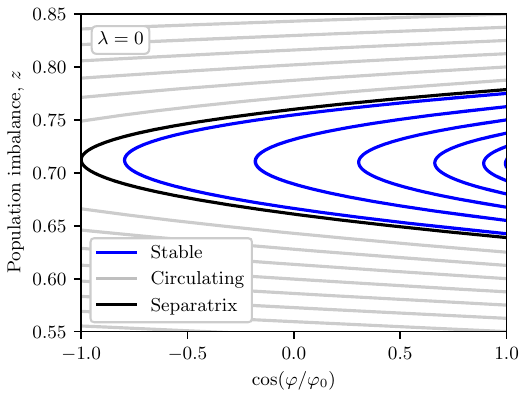}
    \caption{\label{fig:phase-portrait}
    Phase portraits of the asymmetric analog system with and without modulation, with parameters given in Table~\ref{tab:parameters}.
    We obtain each curve by numerically integrating the mean-field equations of motion~\eqref{eq:mean-field-equations-of-motion} from different initial conditions.
    In the left panel, we see a family of trajectories corresponding to small oscillations around the false vacuum (red).
    In the right panel, we see that no such trajectories exist in the absence of modulation.}
\end{figure*}

In Fig.~\ref{fig:phase-portrait} we show phase portraits for the dynamical system defined by Eq.~\eqref{eq:mean-field-equations-of-motion}, with and without modulation.
For modulation of sufficiently large amplitude, $\lambda>1$, we see that both stationary solutions are classically stable, with small perturbations giving rise to bound orbits.
However, the energy of the antiphase state is greater than that of the in-phase state, rendering it metastable once we include vacuum fluctuations.
As expected, we also see that the antiphase state becomes unstable if we switch off the modulation.

\subsection*{Relativistic equations of motion}

We now consider the full field theory, including inhomogeneous modes in both species, in order to establish the relativistic equations of motion~\eqref{eq:rel-eoms}.
It is convenient to write the atomic fields as
    \begin{align}
    \begin{split}
        \psi_\down&=\sqrt{n(1+z)}\exp(f_\down+\rmi\phi_\down),\\
        \psi_\up&=\sqrt{n(1-z)}\exp(f_\up+\rmi\phi_\up),
    \end{split}
    \end{align}
    where $n$ and $z$ are constants that define fixed background densities, and $f_i$ are fields encoding deviations from this background.
We assume that $f_i\sim\order*{\epsilon}$, but do not consider any fixed background values for the phases $\phi_i$; this is crucial to obtain a nonlinear relativistic theory with a false-vacuum potential.
Assuming that field gradients and the detuning are small, $\hbar^2\laplacian/(\kappa nm)\sim\hbar\delta/(\kappa n)\sim\order*{\epsilon}$, we obtain first-order equations of motion to linear order in $\epsilon$ for the total and relative combinations
    \begin{align}
    \begin{split}
        f_+&=(1+z)f_\down+(1-z)f_\up,\quad f_-=f_\down-f_\up,\\
        \phi_+&=(1+z)\phi_\down+(1-z)\phi_\up,\quad\phi_-=\phi_\down-\phi_\up.
    \end{split}
    \end{align}
Taking a second time derivative of the $\phi_\pm$ equations of motion, and plugging in the first-order $f_\pm$ equations of motion, we eliminate $f_\pm$ entirely and obtain second-order equations for the phases,
    \begin{equation}
        \partial_t^2\mqty(\phi_+ \\ \phi_-)=\vb*M\mqty(\laplacian\phi_+ \\ \laplacian\phi_--\frac{4\epsilon\kappa nm}{\hbar^2}\sin\phi_-(1+\lambda^2\cos\phi_-)),
    \end{equation}
    where the two fields are coupled by a mixing matrix,
    \begin{equation}
    \label{eq:mixing-matrix}
        \vb*M=\frac{n}{m}\mqty(2(g-\Delta z)-\kappa(1-z^2) & (\kappa z-\Delta)(1-z^2) \\ \kappa z-\Delta & \kappa(1-z^2)).
    \end{equation}
(Note that there are $\order*{\epsilon}$ terms missing from this expression; capturing those would require us to include $\order*{\epsilon^2}$ contributions in the Hamiltonian.)
We see immediately that for $z=\Delta/\kappa+\order*{\epsilon}$ the off-diagonal elements of this matrix both vanish, giving the decoupled relativistic equations of motion~\eqref{eq:rel-eoms}.
The on-diagonal elements of $\vb*M$ then define the sound speeds of the two fields.

\subsection*{Vacuum fluctuations}

We now consider the statistics of quantum fluctuations in the analog false vacuum, and show that these match those of the corresponding relativistic fields.
This generalizes the results in Ref.~\cite{Jenkins:2023eez}, which first established this equivalence for symmetric mixtures.

We promote the $\psi_i$ from classical fields to quantum operators, and study small fluctuations around the false vacuum,
    \begin{equation}
        \hat{\psi}_\down=\sqrt{n(1+z)}+\updelta\hat{\psi}_\down,\quad\hat{\psi}_\up=-(\sqrt{n(1-z)}+\updelta\hat{\psi}_\up).
    \end{equation}
We work in Fourier space, and define the total and relative operators,
    \begin{align}
    \begin{split}
        \hat{\psi}^+_{\vb*k}=\int_V\frac{\dd{\vb*x}}{\sqrt{2V}}\rme^{-\rmi\vb*k\vdot\vb*x}\qty(\sqrt{1+z}\,\updelta\hat{\psi}_\down(\vb*x)+\sqrt{1-z}\,\updelta\hat{\psi}_\up(\vb*x)),\\
        \hat{\psi}^-_{\vb*k}=\int_V\frac{\dd{\vb*x}}{\sqrt{2V}}\rme^{-\rmi\vb*k\vdot\vb*x}\qty(\sqrt{1-z}\,\updelta\hat{\psi}_\down(\vb*x)-\sqrt{1+z}\,\updelta\hat{\psi}_\up(\vb*x)),
    \end{split}
    \end{align}
    which are normalized so that they obey the standard bosonic commutation relations.
To quadratic order in these fluctuations, the Hamiltonian~\eqref{eq:effective-hamiltonian} decouples into two independent sectors, $\hat{H}_\mathrm{eff}=E_0+\hat{H}_++\hat{H}_-$, if and only if the decoupling condition~\eqref{eq:decoupling} holds.
Each sector is diagonalized by a Bogoliubov transformation,
    \begin{equation}
        \hat{H}_\pm=\sum_{\vb*k\ne\vb*0}\hbar\omega^\pm_k\hat{a}^{\pm\dagger}_{\vb*k}\hat{a}^\pm_{\vb*k},\quad\hat{a}^\pm_{\vb*k}=-\rmi(u^\pm_k\hat{\psi}^\pm_{\vb*k}+v^\pm_k\hat{\psi}^{\pm\dagger}_{-\vb*k}),
    \end{equation}
    where $u^\pm_k$, $v^\pm_k$ are real coefficients satisfying $(u^\pm_k)^2-(v^\pm_k)^2=1$.
We then interpret $\hat{a}^{\pm\dagger}_{\vb*k},\hat{a}^\pm_{\vb*k}$ as creation and annihilation operators for the normal modes of the fields, which have energies $\hbar\omega^\pm_k$.

We find that the Bogoliubov coefficients that achieve this diagonalization are
    \begin{align}
    \begin{split}
        (u^+_k)^2=(v^+_k)^2+1&=\frac{mc_\vartheta^2}{2\hbar\omega^+_k}(\xi_\vartheta^2k^2+1)+\frac{1}{2},\\
        (u^-_k)^2=(v^-_k)^2+1&=\frac{mc_\varphi^2}{2\hbar\omega^-_k}\qty(\xi_\varphi^2k^2+1-\frac{m_0^2}{2m^2})+\frac{1}{2},
    \end{split}
    \end{align}
    and the corresponding dispersion relationships are
    \begin{align}
    \begin{split}
        (\omega_k^+)^2&=c_\vartheta^2k^2\qty(1+\frac{1}{2}\xi_\vartheta^2k^2),\\
        (\omega_k^-)^2&=\qty(c_\varphi^2k^2+\frac{m_\mathrm{fv}^2c_\varphi^4}{\hbar^2})\qty(1+\frac{1}{2}\xi_\varphi^2k^2-\frac{m_0^2}{4m^2}(\lambda^2+1)).
    \end{split}
    \end{align}
On scales much larger than their respective healing lengths, $\xi_i^2k^2\ll1$, these reduce to the expected relativistic expressions (massless and massive, respectively),
    \begin{equation}
        (\omega^+_k)^2\simeq c_\vartheta^2k^2,\qquad(\omega^-_k)^2\simeq c_\varphi^2k^2+\frac{m_\mathrm{fv}^2c_\varphi^4}{\hbar^2}.
    \end{equation}
In the same limit, the $\vartheta$ and $\varphi$ fields themselves can be written in terms of the Bogoliubov modes as
    \begin{equation}
        \hat{\vartheta}_{\vb*k}\simeq\sqrt{\frac{\hbar c_\vartheta^2}{2\omega^+_k}}(\hat{a}^+_{\vb*k}+\hat{a}^{+\dagger}_{-\vb*k}),\qquad\hat{\varphi}_{\vb*k}\simeq\sqrt{\frac{\hbar c_\varphi^2}{2\omega^-_k}}(\hat{a}^-_{\vb*k}+\hat{a}^{-\dagger}_{-\vb*k}),
    \end{equation}
    which match the expressions for the corresponding (canonically-normalized) relativistic fields.
This guarantees that the statistics of the vacuum fluctuations will match those of their relativistic counterparts in the long-wavelength regime.

One effect we have neglected here is the contribution of the zero-point energies of these Bogoliubov modes to the Hamiltonian.
These Lee-Huang-Yang (LHY) terms play a vital role in studies of quantum droplet formation in Bose-Bose mixtures~\cite{Cabrera:2018qld,Semeghini:2018sbq,Cheiney:2018sol,Ferioli:2019col}.
We have checked explicitly that including the resulting corrections to the equations of motion via a local density approximation does not spoil the decoupled pseudorelativistic theory described above.
Instead, the LHY terms give corrections to quantities such as the sound speeds and the critical $z$ for decoupling, which can be computed perturbatively in powers of $\bar{n}^{-1}$.

\subsection*{Euclidean bounce action}

Much of the scientific insight that will be derived from analogue vacuum decay experiments will come from comparing against instanton calculations in Euclidean time $\tau=\rmi t$.
These predict a decay rate per unit volume that can be written as~\cite{Coleman:1977py,Callan:1977pt}
    \begin{equation}
        \frac{\Gamma}{V}\simeq A\,\qty(\frac{B}{2\uppi\hbar})^{(d+1)/2}\exp(-B/\hbar),
    \end{equation}
    where $d$ is the number of spatial dimensions, $A$ is a fluctuation determinant prefactor, and $B$ is the Euclidean `bounce' action describing the nucleation event.
In our effective relativistic theory, this action is
    \begin{equation}
        B=\int\dd{\tau}\int_V\dd{\vb*x}\qty(\frac{1}{2c_\varphi^2}\dot{\varphi}_\mathrm{b}^2+\frac{1}{2}\abs{\grad\varphi_\mathrm{b}}^2+U(\varphi_\mathrm{b})-U(\varphi_\mathrm{fv})),
    \end{equation}
    where $\varphi_\mathrm{b}(\tau,\vb*x)$ is the $O(d+1)$-symmetric bounce solution, and $U(\varphi)$ is the potential~\eqref{eq:potential}.
It is instructive to factor out the dimensionful quantities in this expression, to extract the overall scaling with experimental parameters.
Using our definitions of $m_0$, $c_\varphi$, and $\varphi_0$, we then obtain
    \begin{equation}
        B/\hbar=\frac{\bar{n}_\varphi}{2^{d/2}}\,\epsilon^{(1-d)/2}(1-z^2)^{(1+d)/2}\bar{B},
    \end{equation}
    where $\bar{B}$ is a dimensionless bounce action that depends only on $\lambda$.
As discussed in the main text, the action is proportional to the dimensionless density $\bar{n}_\varphi=n\xi_\varphi^d$, which controls the amplitude of vacuum fluctuations.
We also see that, because $\epsilon\ll1$, the action becomes parametrically larger as the number of dimensions increases, suppressing the decay rate.
(Recall that this is part of our motivation for considering a 1D example setup.)
\end{document}